\documentclass[runningheads]{llncs}
\usepackage{graphicx}
\usepackage{subfigure}
\usepackage{amsmath,bm}
\usepackage{booktabs}
\usepackage{multirow}
\usepackage{multicol}
\usepackage{algorithm}
\usepackage{algorithmic}
\usepackage{multirow} 
\usepackage{array}
\usepackage{bm,bbm}
\usepackage{appendix}
\usepackage{flushend}
\usepackage{xspace}
\usepackage{amssymb}
\usepackage{hyperref}
\usepackage[misc]{ifsym}

\newcommand{\eg}{\emph{e.g.,}\xspace}
\newcommand{\ie}{\emph{i.e.,}\xspace}
\newcommand{\wrt}{\emph{w.r.t.}\xspace}
\newsavebox\CBox
\def\textBF#1{\sbox\CBox{#1}\resizebox{\wd\CBox}{\ht\CBox}{\textbf{#1}}}
\begin{document}
\title{XDM: Improving Sequential Deep Matching with Unclicked User Behaviors for Recommender System}
\titlerunning{XDM}
\author{Fuyu Lv\inst{1}$^{(\textrm{\Letter})}$ \and
Mengxue Li\inst{1} \and
Tonglei Guo\inst{1} \and 
Changlong Yu\inst{2} \and
Fei Sun\inst{1} \and
Taiwei Jin\inst{1} \and
Wilfred Ng\inst{2}}
% \thanks{denotes the corresponding author.}
%
\authorrunning{F. Lv et al.}
\institute{Alibaba Group, Hangzhou, China
\email{\{fuyu.lfy,lydia.lmx,tonglei.gtl,ofey.sf,taiwei.jtw\}}@alibaba-inc.com\\ \and
The Hong Kong University of Science and Technology, Hong Kong, China\\
\email{\{cyuaq,wilfred\}@cse.ust.hk}}
\maketitle              % typeset the header of the contribution
\begin{abstract}
Deep learning-based sequential recommender systems have recently attracted increasing attention from both academia and industry. 
Most of industrial Embedding-Based Retrieval~(EBR) systems for recommendation share the similar ideas with sequential recommenders.
Among them, how to comprehensively capture sequential user interest is a fundamental problem. 
However, most existing sequential recommendation models take as input clicked or purchased behavior sequences from user-item interactions. 
This leads to incomprehensive user representation and sub-optimal model performance, since they ignore the complete user behavior exposure data, \ie items impressed yet unclicked by users.
In this work, we attempt to incorporate and model those unclicked item sequences using a new learning approach in order to explore better sequential recommendation technique.
An efficient triplet metric learning algorithm is proposed to appropriately learn the representation of unclicked items. 
Our method can be simply integrated with existing sequential recommendation models by a confidence fusion network and further gain better user representation. 
The offline experimental results based on real-world E-commerce data demonstrate the effectiveness and verify the importance of unclicked items in sequential recommendation.
Moreover we deploy our new model~(named XDM) into EBR of recommender system at Taobao, outperforming the previous deployed generation SDM.

\keywords{User Behavior Modeling \and Sequential Recommendation \and Metric Learning \and Embedding-based Retrieval.}
\end{abstract}
\section{Introduction}
In order to reduce information overload and satisfy customers' diverse online service needs~(\eg E-commerce, music, and movies), personalized recommender systems~(RS) have become increasingly important. 
Traditional recommendation algorithms~(collaborative filtering~\cite{sarwar2001item} and content-based filtering~\cite{pazzani2007content}) only model users' long-term preference, while ignore dynamic interest in users' behavior sequences. 
Hence sequential recommendation~(SR) is introduced to model sequential user behaviors in history to generate user representation by considering time dependency of user-item interactions.

Moreover, SR sheds light on the rapid development of embedding-based retrieval~(EBR) system~(\textit{a.k.a} deep matching or deep candidate generation) for recommendation in industry~(\eg YouTubeDNN~\cite{covington2016deep}, SDM~\cite{lv2019sdm}, and MIND~\cite{li2019multi}).
The key to EBR system is understanding the evolution of users' preference. 
However, those models as well as most existing SR models~(\eg GRU4REC~\cite{Hidasi:GRU4REC}, NARM~\cite{Li:NARM}, and Caser~\cite{tang2018personalized}) only take as input sequential clicked or purchased behaviors for user modeling. 
They pay little attention to model more abundant exposure data in users' complete behavior sequences, \ie those items that were impressed to users yet not clicked~(refer to \textit{unclicked items} in this paper). 
% The unclicked items make up a relatively significant share of the whole user-item exposure data. SASRec~\cite{Kang:SASRec}
They are of less interest to users, which influence users' future behaviors and can bring better understandings about users’ preference.
Those items also contain valuable signal on users' dynamic preference, which can complement the clicked data. 
Modeling users' preference ignoring the unclicked behavior sequences leads to incomprehensive user representation and limits the capacity and performance of SR. 

\begin{figure}
    \centering
    \includegraphics[scale=0.25]{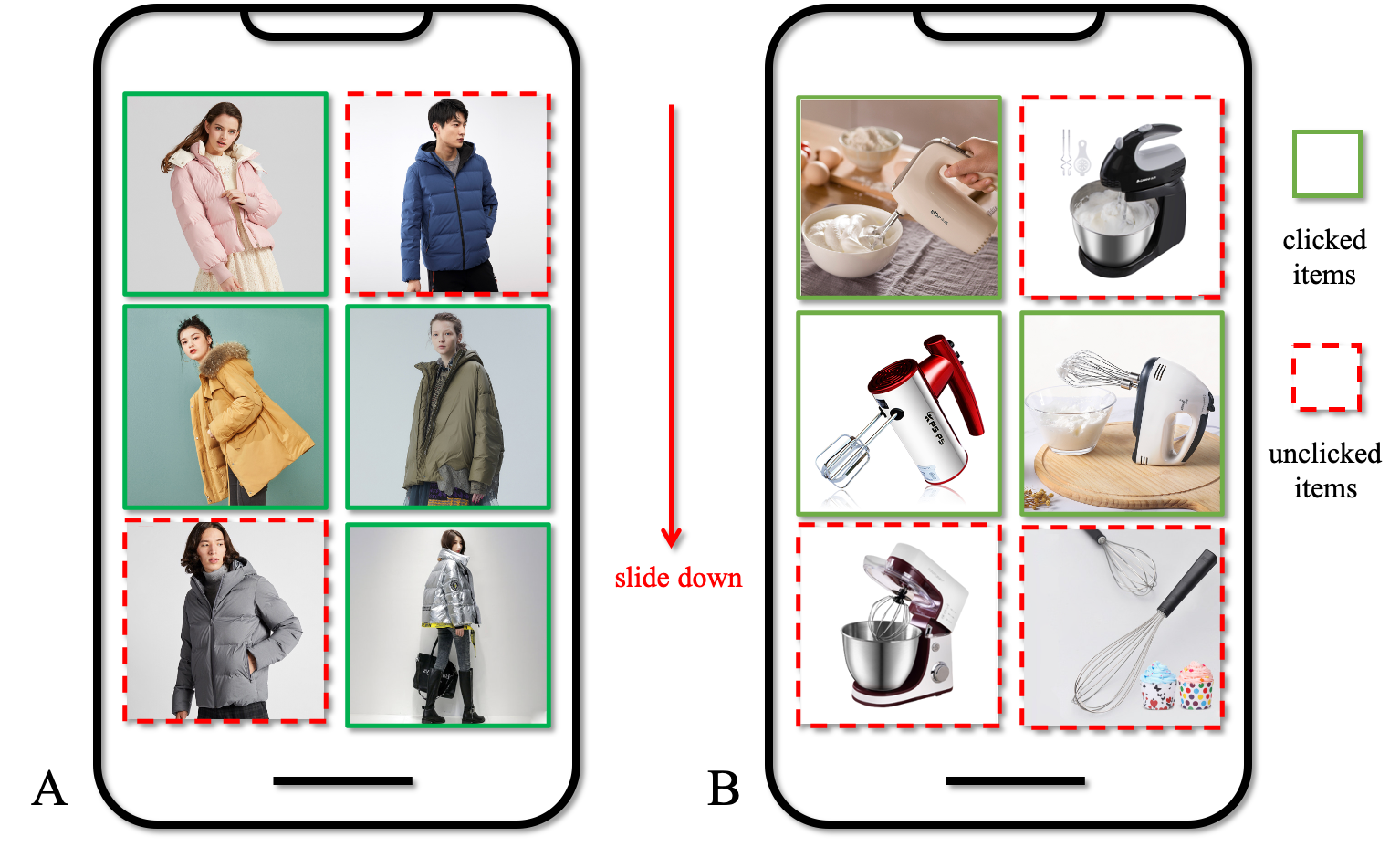}
    \caption{
    % An example of one user's behavior logs from Taobao E-commerce site. 
    Previous sequential recommendations only consider clicked sequences while unclicked sequences are also informative. For example, 
    (A) Users choose women's wear rather than men's wear. (B) Users selectively clicked the same type of products.}
    \label{fig:my_label}
\end{figure}

In this work, we aim to integrate the valuable unclicked item sequences with clicked ones as complete user behaviors into SR models' input to enhance performances of sequential deep matching.
Though it is novel in SR, prior works~\cite{ding2019reinforced,ijcai2020-349,zhao2018recommendations} explore for the general recommendations. 
Compared with SR, they focus on quite different tasks~(\ie matrix factorization~\cite{ding2019reinforced}, reinforcement recommender~\cite{zhao2018recommendations} or click-through rate prediction~\cite{ijcai2020-349}) and specially show different settings such as task definition, training/test sample construction and evaluation.
Besides, their modelings of unclicked sequences remain at the feature level without complex interactions with clicked ones. 
It is believable that clicked and unclicked behaviors affect each other.
Naturally we start to think about effectively incorporating them together from the model level.  % with considering the relationship with clicked ones. ~\cite{wang2018modeling}
Firstly, we derive two important characteristics observed from real-life cases. 
\textbf{1)}~As introduced, it is obvious that unclicked items reflect users' dislikes to some extent compared with clicked ones as shown in Figure~\ref{fig:my_label}(A). 
\textbf{2)}~On the other hand, this kind of items are not those that users particularly dislike compared with a random recommended item. 
The skipped unclicked items can be seen as an \textit{intermediate feedback} between clicked and random recommended items.
Because a modern RS recommends items in which users are probably interested by personalized algorithms.
Users choose to skip items possibly due to many other complex factors, such as price of items displayed nearby, seasonal nature of items or hot consumer trends.
Illustrated in Figure~\ref{fig:my_label}(B), all of these items impressed to users at least partly conform to users’ preference, but the user only select a few of them to click.
Those unclicked items obviously are not random items.

Based on these observations, we propose a new metric learning algorithm to learn the ``intermediate'' representations of unclicked item sequences in SR. 
Specifically, we first project sequential clicked and unclicked behaviors as well as labels into the vector space by deep neural networks~(\eg LSTM, self-attention, and MLP), where Euclidean distance is used as metric measurement. 
The labels are next clicked items after the current user sequence, which represent the true vector of user interest. 
We consider triplet relations among those different vectors: \textbf{1)} clicked and unclicked item vectors, and \textbf{2)} clicked and label item vectors. 
The key idea is to regularize the model by enforcing that the representation of clicked sequence should be far away from the unclicked one. 
Meanwhile the accompanying direction of regularization is applied to clicked and label item representations, which pushes the correct optimization of clicked representation towards the label vector. 
Moreover, the properties of \textit{intermediate feedback} of unclicked items are ensured by adding a predefined margin, which controls the maximum distance between clicked and unclicked vectors.
The clicked and unclicked vectors are combined by a confidence fusion network, which dynamically learns the fusion weight of unclicked items, to get the final user representation.

The offline experimental results based on two real-world E-commerce datasets demonstrate the effectiveness. 
Further experiments have been conducted to understand the importance of unclicked items in the sequential recommendation.
We successfully deploy our new model, named XDM, into EBR of recommender system at Taobao, replacing the previous generation SDM~\cite{lv2019sdm}.
Online experiments demonstrate that XDM leads to improved engagement metrics over SDM.
The main contributions of this paper are summarized below:

\begin{itemize}
    \item We identify the importance of unclicked items in SR and integrate them into models for complete sequential user behavior modeling.
    \item We propose XDM based on triplet metric learning and a confidence fusion network to model users' unclicked together with clicked item sequences. It dynamically controls relationships between different representations to achieve accurate recommendation.
    \item We demonstrate the effectiveness of XDM on real-world E-commerce data for this topic, which would shed light on more research of incorporating unclicked item sequences. Our model has also been successfully deployed on the production environment of recommender system at Taobao.
\end{itemize}

\section{Our Approach}
% In this section, we first introduce the notation used in XDM and formulate our recommendation task. 
% Then we review the base sequential recommendation models. At last, we will introduce and analyze XDM in detail. 
% At last, the triplet loss structure is briefly introduced, which have been widely adopted for deep metric learning.

\subsection{Problem Formulation}
Let $\mathcal{U} = \{u_1, \dots, u_m\}$ denote the set of users, and $\mathcal{I} = \{i_1, \dots, i_n\}$ denote the set of items. % u_2, i_2, 
Our task focuses on implicit recommender systems. %, such as E-commerce. 
% Implicit recommender systems indirectly model users’ preference through behaviors like watching videos, purchasing products, and clicking items. 
For a user $u \in \mathcal{U}$, we record the user's clicking interactions in the ascending order with time $t$ and get the clicked sequence, namely $\mathcal{S}_u^+=\{i^+_1,\dots,i^+_t,\dots,i^+_{n_p}\}$.
% \footnote{We omit the subscript of $u$ from the item indices without loss of clarity.}. 
% For a user $u \in \mathcal{U}$, we record the user's clicking interactions, then sort these records by interaction time $t$ in the ascending order and get the clicked sequence, namely $\mathcal{S}_u^+=\{i^+_1,\dots,i^+_t,\dots,i^+_{n_p}\}$\footnote{We omit the subscript of $u$ from the item indices without loss of clarity.}. 
The unclicked sequence~(items impressed to $u$ yet without clicking interactions) is formed by the same way, namely $\mathcal{S}_u^-=\{i^-_1,\dots,i^-_t,\dots,i^-_{n_n}\}$. The two sequences make up the complete sequential user behaviors $\mathcal{S}_u=\mathcal{S}_u^+ \cup \mathcal{S}_u^-$. 
In fact, clicked and unclicked items appear alternately in the same user sequence.
We partition them into individual sequences to simplify problem definition in our work.
% Their strong dependence in a same sequence will remain for future work.
% We describe each item $i \in \mathcal{I}$ from different feature scales, \ie item ID, leaf category, first level category, brand and shop, which are denoted as side information set $\mathcal{F}$. % to avoid making the problem more complex.
% Based on these preliminaries, we can define the sequential recommendation task. 

Given $\mathcal{S}_{u,t}$, we would like to predict the items set $\mathcal{I}^{pre}_{u,t} \subset \mathcal{I}$ that the user will interact after $t$. % the user $u$'s complete historical behavior sequence
In the process of modeling, all types of user behaviors are encoded into vectors of the same dimension $L_e$. Following~\cite{lv2019sdm}, we take next $k$ clicked items after $\mathcal{S}_{u,t}$ as target items (labels) denoted as $\mathcal{C}_{u,t}=\{c_1,\dots,c_k\}$.
In practice, due to the strict requirement of latency, industrial recommender systems usually consist of two stages, matching and ranking. 
The matching, also called EBR if embedding techniques used, corresponds to retrieving Top-$k$ candidates. %, while the ranking stage is used for sorting candidates by more precise scores. 
Our paper mainly focuses on improving the effectiveness in EBR.

\subsection{Base Sequential Recommendation}\label{sec:method}
% In real industrial recommender systems, users browse and interact with items in chronological order, where the containing items in the same sequence, \ie $\mathcal{S}_u^+$, may be closely relevant. 
% This property facilitates a non-trivial recommendation task: sequential user modeling and successive item recommendation based on users' historical behaviors. 
% Recently deep learning-based models have demonstrated great power in capturing and characterizing the temporal dependency in sequence data. 
% Various effective sequential deep recommenders are proposed, where RNN~\cite{Hidasi:GRU4REC,Quadrana:HRNN}, CNN~(Convolutional Neural Network)~\cite{tang2018personalized}, self-attention~\cite{Liu:STAMP,Kang:SASRec} or memory network~\cite{Pi:MIMN,chen2018sequential} are adopted to model behavior sequences.

% the interaction sequence of user $u$ at time $t$
Given $\mathcal{S}_{u,t}^{+}=\{i^+_1,i^+_2,\dots,i^+_t\}$, a deep sequential recommender computes the user representation vector $\bm{h}_{u,t} \in \mathbb{R}^{L_e}$ as:
\begin{equation}
    \bm{h}_{u,t} = {\rm DSR}(\mathcal{S}^+_{u,t}, \bm{e_u}; \Theta)
\end{equation}
where $\bm{e}_u \in \mathbb{R}^{L_e}$ is the user profile (gender, sex, etc.) embedding. 
$\rm DSR$ means \textbf{D}eep \textbf{S}equential \textbf{R}ecommenders for short. 
$\Theta$ denotes all the model parameters. 
% As for the input of ${\rm DSR}$, 
Each item $i \in \mathcal{S}_{u,t}^{+}$ is mapped into an \textit{item embedding} vector $\bm{q}_i \in \mathbb{R}^{L_e}$.
% , which is called \textit{item embedding} and can be learned or fixed. 
% $\bm{h}_{u,t}$ represents the user's current sequential preferences.
% In addition to the item side, $\rm DSR$ encodes user $u$'s clicked behavior sequence into vector $\bm{h}_{u,t}$, which represents the user's current preferences and hence is named by \textit{sequential preference representation}. 

To generate sequential recommendations for user $u$ at time $t$, we rank a candidate item $i$ by computing the recommendation score $\hat{y}_{u,i,t}$ according to:
\begin{equation}\label{eq:ip}
    \hat{y}_{u,i,t} = g(u, i ,t) = \bm{h}_{u,t}^\mathrm{T} \cdot \bm{q}_i
\end{equation}
where $g(\cdot)$ is the score function, implemented as the inner product between $\bm{h}_{u,t}$ and $\bm{q}_i$. 
After obtaining scores of all items, we can select Top-$k$ items for recommendation. 
As the item candidates of industrial recommender systems are from a very large corpus, the online process of scoring all items is generally replaced with fast K nearest neighbors~(KNN) algorithm.

\subsection{Sequential Recommendation with Unclicked User Behaviors}

The base DSR only take as input $\mathcal{S}_{u,t}^+$ and recommends items according to $\bm{h}_{u,t}$. They ignore the influence of $\mathcal{S}_{u,t}^-$.
We propose to model $\mathcal{S}_{u,t}^-=\{i^-_1,i^-_2,\dots,i^-_t\}$ as a plug-in module on basis of DSR.  
% Starting from a base DSR, the modeling of unclicked items is integrated into the base model for more accurate representation. 
Here we choose \textit{SDM}~\cite{lv2019sdm} as the base DSR due to its capacity of handling with large-scale data for efficient deployed industry applications.

% the following careful considerations: 
% \begin{itemize}
%     \item the state-of-the-art performance of several sequential recommendation benchmarks. 
%     \item capacity of handling with large-scale data for efficient deployed industry applications. 
% \end{itemize}

% \textit{SDM} considers short and long-term user interest when modeling sequential user behaviors. Multiple interest in a user's session is emphasized and modeled by LSTM and Transformer neural structures. 
% The long-term interest is combined with the short one by an efficient gating mechanism, which takes correlation between short and long-term into account. 

\textbf{Metric Learning for Unclicked Items.}\label{sec:metric} 
Compared with clicked ones, unclicked items reflect users' dislikes to some extent, but they are not those users particularly dislike compared to a random recommended item. Because a modern RS recommends items which at least partly conform to users' preference by personalized algorithms.
Thus unclicked items can be intuitively treated as the \textit{intermediate feedback} between clicked and random recommended items.
Therefore, for user $u$'s $\mathcal{S}_{u,t}^-$, it should have an intermediate representation of vector $\bm{n}_{u,t}$ between $\mathcal{S}_{u,t}^+$ and random recommended items. 
% Due to the personalized recommendation needs, we need to capture the specific and detailed preference differences of each user. Thus, it is difficult to find an appropriate representation $\bm{n}_{u,t}$ for different users. 

To solve this problem, we introduce metric learning to control the representation of $\mathcal{S}_{u,t}^-$.
% Inspired by the success of metric learning for the tasks of visual retrieval~\cite{oh2016deep,hermans2017defense} and classification~\cite{qian2015fine}, we formally define the optimization in our recommendation scenarios and extend it with task-specific designs.
% By this means, we are able to differentiate the preference levels between unclicked items and others for modeling users' comprehensive representation in an accurate and flexible way.
Specifically, the first step is to project the sequence $\mathcal{S}_{u,t}^-$ into vector space. 
% With the reminder of \textit{SDM} as our base DSR, 
We encode item $i \in \mathcal{S}_{u,t}^-$ denoted as $\bm{q}_i$, which is the same as $i \in \mathcal{S}_{u,t}^+$. % from different feature scales in $\mathcal{F}$
On account of huge volume of unclicked items, we simply average all the $\bm{q}_i$ in $\mathcal{S}_{u,t}^-$ and then use feed-forward network to generate the embedding of unclicked items $\bm{n}_{u,t} \in \mathbb{R}^{L_e}$, described as:
\begin{equation}
    \bm{n}_{u,t} = f(\frac{1}{|\mathcal{S}_{u,t}^-|} \sum_{i=1}^{|\mathcal{S}_{u,t}^-|} \bm{q}_i)
\end{equation}
where $f(\cdot)$ represents non-linear function implemented by feed-forward network with $\tanh$ activation. More complex neural structures \eg Transformer, remain for future work and are not the major points in this paper.

Given a user $u$, now we have $\bm{h}_{u,t}$, $\bm{n}_{u,t}$, and label representation $\bm{c}_{u, t}$. 
Here $\bm{c}_{u, t}$ generated from $\mathcal{C}_{u,t}$ is embedded in the same way of $\bm{n}_{u,t}$. 
Then we use triplet metric learning to construct triple structures among $\bm{h}_{u,t}$, $\bm{n}_{u,t}$, and $\bm{c}_{u, t}$. 
The optimization goal is to make $\bm{h}_{u,t}$ and $\bm{c}_{u, t}$ closer while to make $\bm{n}_{u,t}$ and $\bm{h}_{u, t}$ far away from each other.
The overall triplet optimization is to minimize:

\begin{equation}
    \label{eq:triplet1}
    \mathcal{L}_{tri}=\sum_{u\in\mathcal{U}}\left [ \left \|\bm{h}_{u,t}-\bm{c}_{u,t}\right \|^2_2 - \left \|\bm{h}_{u,t}-\bm{n}_{u,t}\right \|^2_2+m \right ]_+
\end{equation}
where $\left \| \bm{x} \right \|^2_2=\sum_{i=1}^{n}x_i^2$ denotes the squared $l_2$ norm to measure the distance between vectors and the operator $\left[ \cdot \right]_+ = {\rm max}(0,\cdot)$ denotes the hinge function. 
$m > 0$ is the relaxing parameter constraining the maximum margin distance.

\begin{figure}[t!]
  \centering
  \includegraphics[scale=0.08]{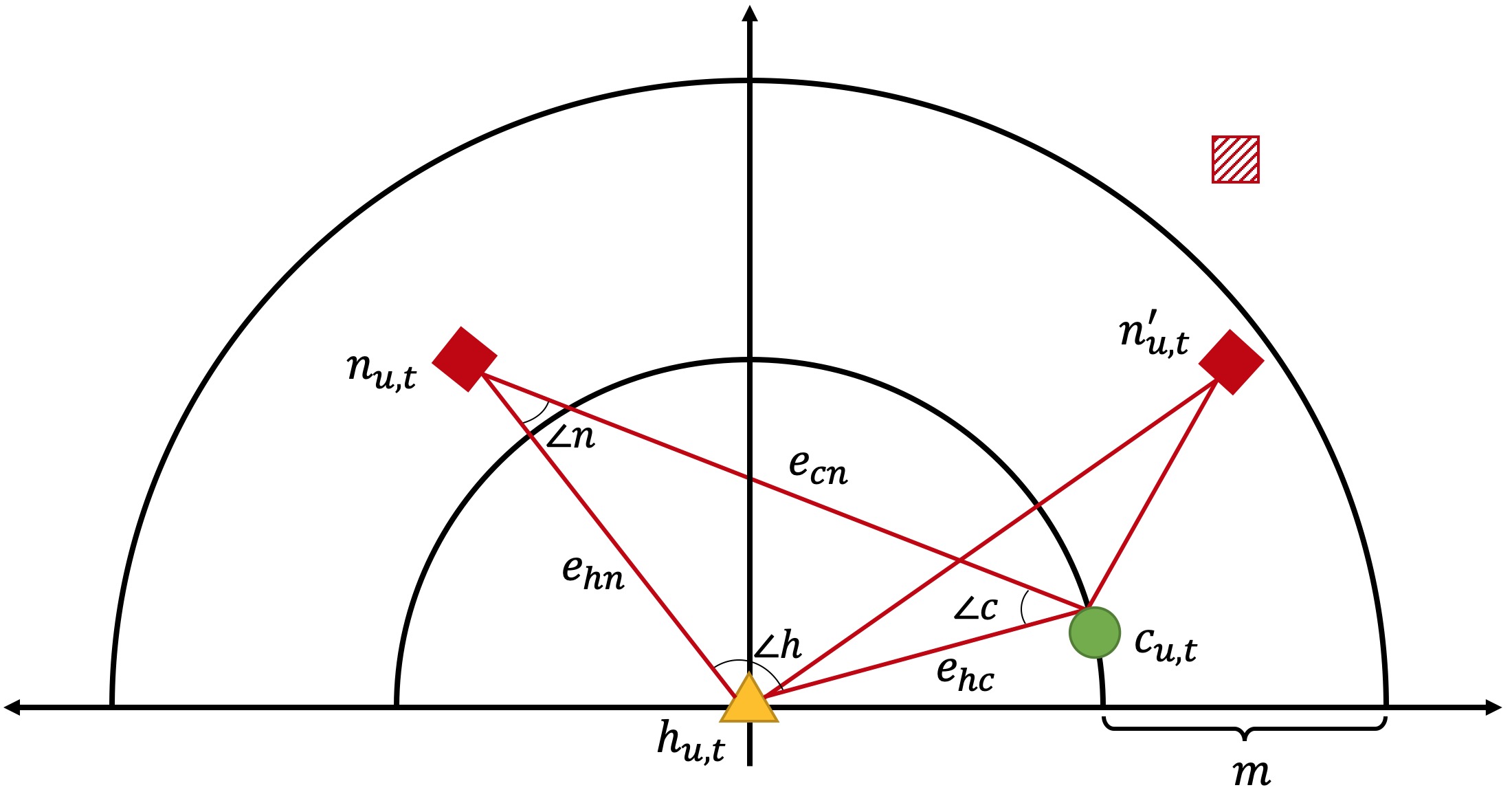}
  \caption{Triplet structure. Red hollow squares represent embedding that have met the constraints and no longer need to be optimized.}
  \label{fig:metric}
\end{figure}

% To understand the role of triplet measurement in the feature space, 
We use an example from two-dimensional space to explain the intuition shown in Figure~\ref{fig:metric}. 
% The triplet $\mathcal{T}=<\bm{h}_{u,t}, \bm{n}_{u,t}, \bm{c}_{u,t}>$ forms a triangle $\triangle hcn$, whose edges are $e_{hn} =\left \|  \bm{h}_{u,t}-\bm{n}_{u,t}\right \|^2_2$, $ e_{cn} =\left \| \bm{c}_{u,t}-\bm{n}_{u,t}\right \|^2_2$, and $e_{hc}=\left \| \bm{h}_{u,t}-\bm{c}_{u,t}\right \|^2_2$, respectively. 
The triplet loss penalizes the shorter edge $e_{hn}$, so that difference between $\bm{h}_{u,t}$ and $\bm{n}_{u,t}$ are significantly large. 
While it will reward the shorter edge $e_{hc}$ to make $\bm{h}_{u,t}$ more similar to $\bm{c}_{u,t}$. 
By introducing margin $m$, we control the maximum difference between $e_{hn}$ and $e_{hc}$ by enforcing $e_{hc} + m \leq e_{hn}$. 
% Hence distinction between $\bm{h}_{u,t}$ and $\bm{n}_{u,t}$ is constrained within $m$. 
It keeps the \textit{intermediate feedback} property of unclicked items between clicked and random recommended items.
The introduction of hinge function is to avoid the further correction of those ``qualified'' triplets.

However, we find that current optimization may lead to undesirable situations, as shown in the Figure~\ref{fig:metric}. 
The movement of $\bm{n}_{u,t}$ to $\bm{n}_{u,t}^{'}$ meets the optimization in Equation~\ref{eq:triplet1}, but $\bm{n}_{u,t}^{'}$ is closer to the $\bm{c}_{u,t}$, which leads to weak distinction between clicked and unclicked item representations. 
In order to eliminate the effect, we derive a symmetrical triplet constraint by increasing $e_{hn}$ and $e_{cn}$ at the same time, \ie adding constraint term $ e_{hc} + m' \leq e_{cn}$. 
Symmetrical constraints are incorporated into Equation~\ref{eq:triplet1} and the new optimization objective is defined as: 

\begin{equation}
    \label{eq:triplet2}
    \begin{aligned}
    \mathcal{L}_{tri}=&\sum_{u\in\mathcal{U}}\left [ \left \|\bm{h}_{u,t}-\bm{c}_{u,t}\right \|^2_2 - \left \|\bm{h}_{u,t}-\bm{n}_{u,t}\right \|^2_2+m \right ]_+ +\\
    &\sum_{u\in\mathcal{U}}\left [ \left \|\bm{h}_{u,t}-\bm{c}_{u,t}\right \|^2_2 - \left \|\bm{c}_{u,t}-\bm{n}_{u,t}\right \|^2_2+m' \right ]_+\\
    =&\sum_{u\in\mathcal{U}}\!\Big [ 2\!\left \|\bm{h}_{u,t}{-}\bm{c}_{u,t}\right \|^2_2 \!\!{-} \left \|\bm{h}_{u,t}{-}\bm{n}_{u,t}\right \|^2_2\!\!{-}\left \|\bm{c}_{u,t}{-}\bm{n}_{u,t}\right \|^2_2\!\! + m^* \Big]_+
    \end{aligned}
\end{equation}
Here we use $m^*$ to represent the addition of two margins in symmetrical losses.

\textbf{Fusion Network.}\label{sec:fusion}
% \textbf{Feature Fusion.} 
To make better use of unclicked sequences, we attempt to explicitly combine $\bm{n}_{u,t}$ with base DSR. 
We first come up a simple method which directly adopts the difference between $\bm{n}_{u,t}$ and $\bm{h}_{u,t}$. 
% The fusion operation of $\bm{n}_{u,t}$ and $\bm{h}_{u,t}$ is performed on the top of the model. 
The final representation $\bm{z}_{u,t}$ could be formulated as: 
\begin{equation}
    \label{eq:fusion}
    \bm{z}_{u,t} = \bm{h}_{u,t} - \bm{n}_{u,t}
\end{equation}
% However, $\bm{z}_{u,t}$ does not contain any complex feature interaction for better modeling. 
Further we elaborately design a confidence neural network as an activation unit in the fusion process: 
% \begin{equation}
%     \bm{G}_{u,t} = \bm{W}{\rm concat}([\bm{h}_{u,t}, \bm{n}_{u,t}]) + \bm{b}
% \end{equation}
% sequential preference representation negative feedback embedding
% The refined user representation $\bm{\hat{z}}_{u,t}$ is generated by subtracting weighted negative information from original sequential representation: %  of user $u$ at time $t$
\begin{equation}
\begin{aligned}
    \label{eq:fusion_gate}
    \bm{G}_{u,t} &= \sigma(\bm{W}{\rm concat}([\bm{h}_{u,t}, \bm{n}_{u,t}]) + \bm{b}) \\
    \bm{\hat{z}}_{u,t} &= \bm{h}_{u,t} - \bm{G}_{u,t}\odot \bm{n}_{u,t}
\end{aligned}
\end{equation}
where $\bm{G}_{u,t} \in \mathbb{R}^{L_e} $ is used to determine the weight, which indicates how to dynamically combine $\bm{h}_{u,t}$ and $\bm{n}_{u,t}$. $\odot$ is element-wise multiplication. $\bm{W}$ is the weight matrix and $\sigma$ is sigmoid function.

\textbf{Overall Structure.} 
\begin{figure}[t!]
  \centering
  \includegraphics[scale=0.12]{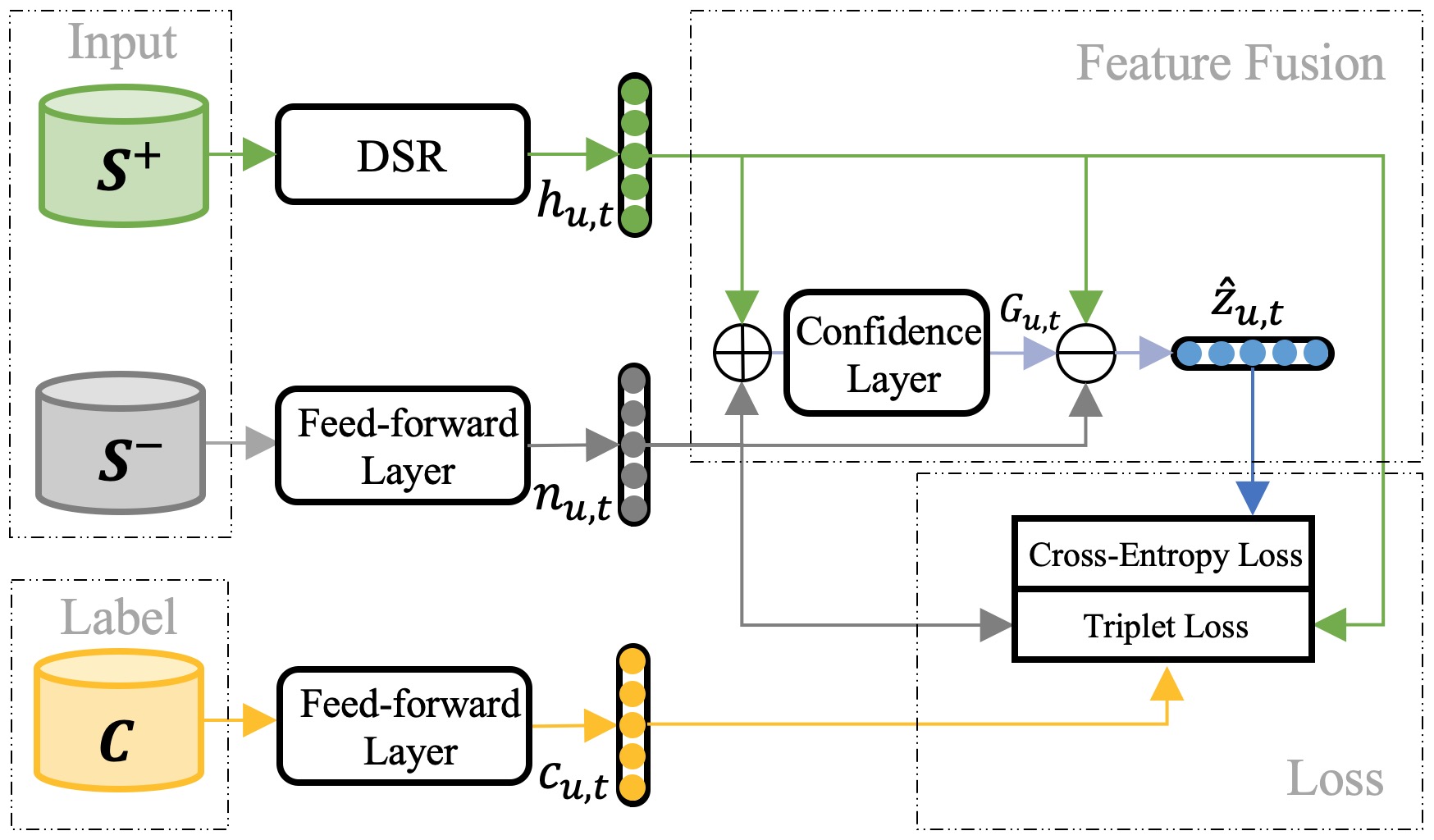}
  \caption{Network structure.}
%   Three types of data (in different colors) are mapped into embedding vectors with the same embedding size. We create the triplet loss based on these embedding vectors and combine the cross-entropy as the overall loss for model optimization.}
  \label{fig:network}
\end{figure}
Figure~\ref{fig:network} illustrates the model structure. Different colors represent different data resources, \ie clicked item sequences, unclicked item sequences and label data. %Boxes represent the network structure and computing units, whose internal parameters are all included in the backward propagation. 
The representation of $\mathcal{S}_{u,t}^+$ and $\mathcal{S}_{u,t}^-$ are concatenated ($\oplus$) as the input of the confidence network. 
Then the confidence network outputs the activation unit $\bm{G}_{u, t}$ for feature fusion.
The recommendations are made based on the final user representation $\bm{\hat{z}}_{u,t}$. 
The optimization of triplet metric learning for unclicked sequences is added to the final loss function.

% \noindent 
\textbf{Loss Function.} 
Besides the triplet loss $\mathcal{L}_{tri}$, we use the sampled-softmax~\cite{jean2014using} method to calculate the cross-entropy loss $\mathcal{L}_{ce}$ over the large amount of items in our real-world dataset for the sake of efficiency. 
Importance sampling~(\eg log-uniform sampler \wrt items frequencies) are conducted to obtain $j$ random negative samples $\mathcal{C}_{u,t}^-$ from unobserved item set $\mathcal{I}/\mathcal{C}_{u,t}$ as most DSR models do~\cite{lv2019sdm}.
The model performs joint optimization according to the overall loss defined as follows:
\begin{equation}
    \begin{aligned}
        &\mathcal{L}_{\rm XDM}= \mathcal{L}_{ce} + \lambda \mathcal{L}_{tri} = \\ \sum_{u\in\mathcal{U}}&{\rm CrossEntropy}\big({\mathcal{C}_{u,t}, {\rm SampledSoftmax}(\bm{\hat{z}}_{u,t}, \mathcal{C}_{u,t}, \mathcal{C}_{u,t}^-)}\big) \ + \\
        \lambda
        \sum_{u\in\mathcal{U}}&\Big [ 2\!\left \|\bm{h}_{u,t}{-}\bm{c}_{u,t}\right \|^2_2 \!\!{-} \left \|\bm{h}_{u,t}{-}\bm{n}_{u,t}\right \|^2_2\!\!{-}\left \|\bm{c}_{u,t}{-}\bm{n}_{u,t}\right \|^2_2\!\!+m^* \Big]_+
    \end{aligned}
\end{equation}
where $\mathcal{C}_{u,t}$ is the positive labels from real behaviors of user $u$ after time $t$. %and $k=|\mathcal{C}_{u,t}|$ is the number of positive samples
The sampled-softmax takes final user representation, positive and negative samples as input, which outputs the prediction probability distribution over items in $\mathcal{C}_{u,t}$. 
% It has the same motivation with large items candidates as the large vocabulary in language modeling~\cite{jean2014using}. 
$\lambda$ is the trade-off coefficient of two loss terms.

\section{EXPERIMENTAL SETUP}

\subsection{Datasets}
As we have discussed, incorporating unclicked sequences into sequential recommendation is a novel exploration, where few of benchmark datasets exist. 
Hence we construct two large-scale datasets collected from the logs of running recommender systems from Mobile Taobao and Tmall platforms\footnote{Popular E-commerce websites with ten millions of active items~(\url{www.taobao.com} and \url{www.tmall.com}).}, within the time period from 2019/12/27 to 2020/01/03. 
The collected data contains user portrait features, user complete behavior sequences including clicked and unclicked items. 
Note that dataset in \cite{lv2019sdm} is also sampled from Taobao, but they do not contain unclicked items and the data is not public available. 
For the training/validation/test dataset split and evaluation pipeline, we directly followed the well-defined procedure in~\cite{lv2019sdm}.

\begin{table}[t]
\small
    \caption{Statistics of experimental datasets}
    \setlength\tabcolsep{3.0pt}
    \centering
    \label{tab:dataset}
    \begin{tabular}{l|l|l}
        \toprule
        Type          & Taobao Dataset     & Tmall Dataset \\
        \midrule
        Num. of Users      & 358,978     & 446,464 \\
        Num. of Items      & 1,078,723   & 908,214 \\
        Num. of Train Data      & 1,041,094   & 1,229,271 \\
        Num. of Test Data       & 17,048      & 20,134 \\
        Avg Short-term Clicked Seq Len & 13.49 & 12.30 \\
        Avg Long-term Clicked Seq Len  & 20.87 & 18.74 \\
        Avg Unclicked Seq Len        & 71.32 & 64.26 \\
        \bottomrule
    \end{tabular}
\end{table}

% During data collection, one crucial issue is to prepare unclicked data, which is a kind of implicit feedback in our experimental platforms. 
% They contain a large number of noises or uninformative signals, while they are the vast majority of all user behaviors. 
% So we propose to use the following pre-defined filtering rules for selecting unclicked data with less noises

% \begin{enumerate}
%     \item Select latest impressed items of a user that were not clicked within the past three days as unclicked candidates.
%     \item Only keep those unclicked items that are exposed to a user more than $k$ times (set $k=1$ in this work) in the above unclicked candidates.
% \end{enumerate}

% Items impressed to a user many times yet not clicked indicate that his/her preference at this stage is not strong. 
% We choose those items naturally to form the unclicked sequence for each user. 
% The maximum length of unclicked sequences was set to 100 for the sake of effective sequential modeling.
% Table~\ref{tab:dataset} shows the statistical information of our datasets.

\subsection{Compared Methods}\label{sec:baselines}
We used the following state-of-the-art sequential recommenders to compare with XDM: \textbf{DNN}~\cite{covington2016deep}, \textbf{GRU4REC}~\cite{Hidasi:GRU4REC}, \textbf{NARM}~\cite{Li:NARM}, \textbf{SHAN}~\cite{ying2018sequential}, \textbf{BINN}~\cite{li2018learning}, and \textbf{SDM}~\cite{lv2019sdm}.
% \begin{itemize}%[leftmargin=*]
%     \item \textbf{DNN}~\cite{covington2016deep}. A classic recommendation method proposed for YouTube based on the deep neural network.
%     \item \textbf{GRU4REC}~\cite{Hidasi:GRU4REC}. It firstly adopts the RNN in the session-based recommender system.
%     \item \textbf{NARM}~\cite{Li:NARM}. Based on GRU4REC, it adds a global and local attention-based module. 
%     \item \textbf{SHAN}~\cite{ying2018sequential}. SHAN combines users’ historical preference with the current shopping demand by a hierarchical attention network.
%     \item \textbf{BINN}~\cite{li2018learning}. BINN applies RNN to encode user's sessions, and connects the current consumption motivation and historical behaviors as a unified user representation.
%     \item \textbf{SDM}~\cite{lv2019sdm}. SDM represents the user behaviors from different levels, which combines short-term sessions and long-term preferences via a gated fusion network.
% \end{itemize}
% \textbf{Our methods.} 
We conducted ablation experiments by gradually adding our proposed modules and compared with the baseline models above.
We employ \textit{SDM} (the best baseline) as the base DSR for modeling clicked sequences.
We name several XDM variants with abbreviated terms. 

\begin{itemize}
    \item \textbf{XDM}. Proposed algorithm of this paper includes both symmetric triplet metric learning~(Equation~\ref{eq:triplet2}) and confidence fusion network~(Equation~\ref{eq:fusion_gate}).
    \item \textbf{XDM~(w/o sym)}. The only difference with XDM is using asymmetric triplet metric learning algorithm~(Equation~\ref{eq:triplet1}). % XDM~(w/o sym)
    \item \textbf{XDM~(w/o fusion+sym)}. XDM only employs asymmetric triplet metric learning algorithm~(Equation~\ref{eq:triplet1}) without any explicit feature fusion. % XDM~(w/o fusion+sym)
    \item \textbf{XDM~(w/o metric)}. XDM only combines unclicked sequences via the fusion network (Equation~\ref{eq:fusion_gate}) to improve feature fusion without metric learning. % XDM~(w/o metric)
    \item \textbf{XDM~(w/o conf+metric)}. XDM only combines unclicked sequences via simple feature fusion~(Equation~\ref{eq:fusion}) without metric learning. % XDM~(w/o conf+metric)
\end{itemize}
Although models in \cite{ijcai2020-349,zhao2018recommendations} are applied in other tasks, they also use unclicked sequences. But their methods are similar to \textit{XDM~(w/o conf+metric)}, which simply regard unclicked sequences as features of neural networks. 
Hence we do not involve them as the baselines for fair comparisons. 

% \noindent
\textbf{Evaluation Metrics.}
% \subsection{Evaluation Metrics}
To evaluate the effectiveness of different methods, we use HR~(Hit Ratio), MRR~(Mean Reciprocal Rank), R~(Recall), and F$_1$ metrics for the Top-$k$ recommendation results, which are also widely used in the previous works~\cite{lv2019sdm,Hidasi:GRU4REC,tang2018personalized}. 
We chose $k= \{50, 80\}$ to report the Top-$k$ performance as \cite{gao2019learning}.
The reason for setting larger $k$ is the huge number of item set $\mathcal{I}$ in our datasets and results over smaller $k$ have larger variance thus uncomparable for the matching stage.
We calculated averaged metrics for the test sets.

\subsection{Implementation Details}\label{sec:settings}
We used the distributed Tensorflow\footnote{\url{https://www.tensorflow.org/guide/distributed_training}} to implement all the methods. 
% The training/test datasets were shared among all the models as well as item and user features.
Results of the baselines and our models on test datasets are reported according to optimal hyper-parameters tuned on validation data.
We used 2 parameter severs~(PSs) and 5 GPU~(Tesla P100-pcie-16GB) workers with average 30 global steps per second to conduct training and inference. 
The embedding size $L_e=128$.
For training, the learning rate was set to 0.1 and the sequences with similar length were selected in a mini-batch whose size is set to 256. 
Adagrad was used as the optimizer and the gradient clipping technique was also adopted.
The next $k = 5$ clicked items after a sequence were taken as the label items in $\mathcal{C}_{u,t}$ in our experiments. The sampled-softmax used $j = 20,000$ random negative samples.
All input feature representation and model parameters were initialized randomly.
For parameters of XDM, we set the margin parameter $m^*$ in the triplet loss to $5$, the trade-off parameter $\lambda$ between cross-entropy loss and triplet loss to $10$. These two parameters were the best results obtained by parameter selection experiment. 
% All types of behaviors were embedded to vectors with the same size $L_e=128$. 
% A single layer feed-forward network with sigmoid activation function was used as the structure of the confidence network at the feature fusion process. 

\section{EXPERIMENT ANALYSIS}

\subsection{Overall Performances}
The experimental results are reported in Table~\ref{tab:offexp} as well as the relative improvement based on the best baseline model. 
DNN performs worst since the average pooling operation ignores the inheritance correlation between items. The performance of GRU4REC and NARM are far beyond the original DNN by modeling the evolution of short-term behavior. Compared to GRU4REC, SHAN and BINN encode more personalized user information, which are significantly better than GRU4REC and beat NARM. 
\textit{SDM} performs well due to jointly modeling long-term and short-term behavior. Also it simulates multiple interests in users' short-term session and combine the long-term preference using a gating network. \textit{XDM} takes \textit{SDM} as the base model. Two modules \ie confidence fusion network and symmetric triplet metric learning, are added to the base model. Results of all metrics are substantially improved. 
\textit{XDM} outperforms it by \textbf{6.21\%} in MRR@50 and \textbf{5.63\%} in $F_1$@50 on the Taobao dataset. Similar trends are also observed on the Tmall dataset. 
This confirms the effectiveness of overall proposed method. 

\begin{table*}%[htb]
    \caption{Top-$k$ recommendation comparison of  different methods. The relative improvements compared to the best baseline~(SDM) are appended on the right starting 
    with ``+/-''. ($k$ is set to 50, 80). * indicates significant improvement of XDM over the baselines in Section~\ref{sec:baselines}. ($p<0.05$ in two-tailed paired t-test).}
    \label{tab:offexp}
    \resizebox{\textwidth}{!}{
    \begin{tabular}{l|c c| c c| c c| c c|c c| c c| c c |c c}
        \toprule%[1.2pt]
        \multicolumn{17}{c}{\qquad \qquad \qquad \textbf{Taobao Dataset}}\\ 
        \hline% \midrule[1.2pt]
        Methods     & \multicolumn{2}{c|}{\textBF{HR@50}} & \multicolumn{2}{c|}{\textBF{MRR@50}} &
        \multicolumn{2}{c|}{\textBF{R@50}} & \multicolumn{2}{c|}{\textBF{F$_1$@50}} & \multicolumn{2}{c|}{\textBF{HR@80}} & \multicolumn{2}{c|}{\textBF{MRR@80}} &
        \multicolumn{2}{c|}{\textBF{R@80}} & \multicolumn{2}{c}{\textBF{F$_1$@80}} \\ 
        \hline% \midrule[1.2pt]
        DNN          & 29.95\% & -17.42\% & 6.65\% & -24.94\% & 1.44\% & -20.44\% & 1.14\% & -19.72\% & 36.74\% & -15.93\% & 6.99\% & -24.43\% & 2.01\% & -18.29\% & 1.19\% & -17.93\% \\
        GRU4REC      & 32.39\% & -10.70\% & 7.85\% & -11.40\% & 1.62\% & -10.50\% & 1.26\% & -11.27\% & 39.25\% & -10.18\% & 8.20\% & -11.35\% & 2.22\% & -9.76\% & 1.30\% & -10.34\% \\
        NARM         & 32.68\% & -9.90\% & 8.20\% & -7.45\% & 1.66\% & -8.29\% & 1.30\% & -8.45\% & 40.27\% & -7.85\% & 8.52\% & -7.89\% & 2.29\% & -6.91\% & 1.34\% & -7.59\% \\
        SHAN         & 34.00\% & -6.26\% & 8.84\% & -0.23\% & 1.81\% & -0.00\% & 1.40\% & -1.41\% & 40.93\% & -6.34\% & 9.20\% & -0.54\% & 2.45\% & -0.41\% & 1.41\% & -2.76\% \\
        BINN         & 36.24\% & -0.08\% & 8.70\% & -1.81\% & 1.73\% & -4.42\% & 1.38\% & -2.82\% & 43.30\% & -0.92\% & 8.64\% & -6.59\% & 2.34\% & -4.88\% & 1.39\% & -4.14\% \\
        SDM          & 36.27\% & - & 8.86\% &- & 1.81\% & - & 1.42\% &- & 43.70\% &- & 9.25\% &- & 2.46\% &- & 1.45\% & - \\
        \hline
         XDM &\textbf{37.97\%}* &	+\textbf{4.69\%} &	\textbf{9.41\%}* &	+\textbf{6.21\%} &	\textbf{1.92\%}* &	+\textbf{6.08\%} &	\textbf{1.50\%}* &	+\textbf{5.63\%} &	\textbf{45.44\%} &	+\textbf{3.98\%} &	\textbf{9.75\%}* &	+\textbf{5.41\%} &	\textbf{2.61\%}* &	+\textbf{6.10\%} &	\textbf{1.53\%}* & +\textbf{5.52\%}\\
        - w/o conf+metric &36.77\% &	+1.38\% &	9.12\% &	+2.93\% &	1.82\% &	+0.55\% &	1.43\% &	+0.70\% &	44.43\% &	+1.67\% &	9.45\% &	+2.16\% &	2.48\% &	+0.81\% &	1.46\% &	+0.69\% \\
         - w/o metric &37.07\% &	+2.21\% &	9.12\% &	+2.93\% &	1.84\% &	+1.66\% &	1.45\% &	+2.11\% &	44.80\% &	+2.52\% &	9.53\% &	+3.03\% &	2.54\% &	+3.25\% &	1.49\% &	+2.76\% \\
        - w/o fusion+sym  &37.13\% &	+2.37\% &	9.09\% &	+2.60\% &	1.89\% &	+4.42\% &	1.47\% &	+3.52\% &	44.87\% &	+2.68\% &	9.51\% &	+2.81\% &	2.58\% &	+4.88\% &	1.52\% & +4.83\%\\
        - w/o sym &37.37\% &	+3.03\% &	9.29\% &	+4.85\% &	1.87\% &	+3.31\% &	1.47\% &	+3.52\% &	45.28\% &	+3.62\% &	9.71\% &	+4.97\% &	2.58\% &	+4.88\% &	1.52\% & +4.83\% \\
       
        \bottomrule
    \end{tabular}}
    \resizebox{\textwidth}{!}{
    \begin{tabular}{l|c c| c c| c c| c c|c c| c c| c c |c c}
        \toprule
         \multicolumn{17}{c}{\qquad \qquad \qquad \textbf{Tmall Dataset}}\\ 
        \hline
        Methods     & \multicolumn{2}{c|}{\textBF{HR@50}} & \multicolumn{2}{c|}{\textBF{MRR@50}} &
        \multicolumn{2}{c|}{\textBF{R@50}} & \multicolumn{2}{c|}{\textBF{F$_1$@50}} & \multicolumn{2}{c|}{\textBF{HR@80}} & \multicolumn{2}{c|}{\textBF{MRR@80}} &
        \multicolumn{2}{c|}{\textBF{R@80}} & \multicolumn{2}{c}{\textBF{F$_1$@80}} \\ 
        \hline
        DNN          & 30.45\% & -19.47\% & 7.01\% & -26.37\% & 1.68\% & -21.86\% & 1.27\% & -21.12\% & 37.64\% & -16.71\% & 7.38\% & -25.23\% & 2.36\% & -18.34\% & 1.33\% & -17.39\% \\
        GRU4REC      & 34.38\% & -9.07\% & 8.94\% & -6.09\% & 1.98\% & -7.91\% & 1.49\% & -7.45\% & 41.17\% & -8.90\% & 9.14\% & -7.40\% & 2.65\% & -8.30\% & 1.48\% & -8.07\% \\
        NARM         & 34.75\% & -8.09\% & 8.96\% & -5.88\% & 2.03\% & -5.58\% & 1.52\% & -5.59\% & 41.89\% & -7.30\% & 9.30\% & -5.78\% & 2.74\% & -5.19\% & 1.52\% & -5.59\% \\
        SHAN         & 35.12\% & -7.11\% & 9.48\% & -0.42\% & 2.15\% & -0.00\% & 1.59\% & -1.24\% & 42.29\% & -6.42\% & 9.87\% & -0.00\% & 2.83\% & -2.08\% & 1.55\% & -3.73\% \\
        BINN         & 37.20\% & -1.61\% & 9.17\% & -3.68\% & 2.04\% & -5.12\% & 1.54\% & -4.35\% & 45.10\% & -0.20\% & 9.65\% & -2.23\% & 2.83\% & -2.08\% & 1.59\% & -1.24\% \\
        SDM          & 37.81\% & - & 9.52\% & - & 2.15\% & - & 1.61\% & - & 45.19\% & - & 9.87\% & - & 2.89\% & - & 1.61\% &-\\
        \hline
         XDM &\textbf{38.91\%}* &	+\textbf{2.91\%} &	\textbf{9.89\%}* &	+\textbf{3.89\%} &	\textbf{2.21\%}* &	+\textbf{2.79\%} &	\textbf{1.66\%}* &	+\textbf{3.11\%} &	{46.57\%}* &	+3.05\% &	10.21\% &	+3.44\% &	\textbf{3.04\%}* &	+\textbf{5.19\%} &	\textbf{1.69\%}* &  +\textbf{4.97\%}\\
        - w/o conf+metric &38.22\% &	+1.08\% &	9.75\% &	+2.42\% &	2.19\% &	+1.86\% &	1.64\% &	+1.86\% &	45.80\% &	+1.35\% &	10.18\% &	+3.14\% &	2.97\% &	+2.77\% &	1.66\% &	+3.11\% \\
        - w/o metric &38.56\% &	+1.98\% &	9.67\% &	+1.58\% &	2.20\% &	+2.33\% &	1.64\% &	+1.86\% &	46.39\% &	+2.66\% &	10.11\% &	+2.43\% &	3.00\% &	+3.81\% &	1.67\% &	+3.73\% \\
        - w/o fusion+sym &38.69\% &	+2.33\% &	9.81\% &	+3.05\% &	\textbf{2.21\%} &	+\textbf{2.79\%} &	1.65\% &	+2.48\% &	46.37\% &	+2.61\% &	\textbf{10.25\%} &	+\textbf{3.85\%} &	3.00\% &	+3.81\% &	1.67\% &  +3.73\%\\
        - w/o sym &38.80\% &	+2.62\% &	9.81\% &	+3.05\% &	\textbf{2.21\%} &	+\textbf{2.79\%} &	\textbf{1.66\%} &	+\textbf{3.11\%} &	\textbf{46.81\%} &	+\textbf{3.58\%} &	\textbf{10.25\%} &	+\textbf{3.85\%} &	3.02\% &	+4.50\% &	1.68\% &  +4.35\%\\
       
        \bottomrule
    \end{tabular}}
\end{table*}

\subsection{Ablation Analysis}\label{sec:ablexp}
To disentangle the capability of each module, we further conducted ablation study and results are also shown in Table~\ref{tab:offexp}. 
\textit{XDM~(w/o conf+metric)} attempts to eliminate noises contained in clicked sequences by using unclicked representation directly, as shown in the Equation~\ref{eq:fusion}. 
The results show that all indicators of this method are slightly improved compared with \textit{SDM}. 
\textit{XDM~(w/o metric)} introduces a confidence network and applies it to weight the unclicked representation in feature fusion process. Results show that almost indicators increase by about 2\%$\sim$3\% on average compared with the base model \textit{SDM} on two datasets. 
These two experiments demonstrate that the unclicked items does reflect negative interest of users, and it plays an important role of denoising in user preference modeling, though equipped with DSR for clicked sequences.

\textit{XDM~(w/o fusion+sym)} only adds an asymmetric triplet loss shown in Equation~\ref{eq:triplet1} without explicit feature fusion operation. The result is positive. 
It reveals the effect of metric learning.
We considered the combination of two modules (confidence fusion and asymmetric metric learning) stated above denoted as \textit{XDM~(w/o sym)}. 
The average results increase about 3\%$\sim$4\% in almost indicators of two dataset, which indicates that the combination of metric learning and feature fusion can make better use of unclicked data. 
Metric learning provides higher-quality representations of unclicked sequences for the feature fusion network.
Along this way, \textit{XDM} further added symmetry constraints, as shown in the Equation~\ref{eq:triplet2}. The average results show it achieves the highest improvement over \textit{SDM} and \textit{XDM~(w/o sym)} in almost evaluations. 
This result shows that symmetric constraints are very important for model learning. 
From comparison results of all variants, we can conclude that significant improvement is produced by the introduction of the confidence network and the triplet metric learning with symmetric constraint. Further, the analysis of metric learning as one of the important modules is included in the Appendix.

% \subsection{Role of Metric Learning}
% Comparing the result of \textit{XDM~(w/o metric)} with \textit{XDM~(w/o sym)} and \textit{XDM}, we can observe that the introduction of metric learning module improves the performance of \textit{XDM} significantly. 
% In this section, we will analyze it in detail. 
% Metric learning algorithm is proposed to model unclicked sequences more appropriately with considering its relationship with clicked sequences. 
% The predefined margin $m^*$ comes into play in a way of threshold limit, and the hyper-parameter $\lambda$ controls the importance ratio between different loss terms. 
%The analysis is divided into two aspects: control margin effect and metric structure effect.
% Note that the goal of experiments below is to show how model performance is affected by these experimental variables. 
% It does not mean we optimize our model on test dataset, though results are reported on test.

% \subsubsection{The effect of margin}
\subsection{The Effect of Margin}
\begin{figure}[t!]
    \centering
    \subfigure[Effect of $m^*$ on HR$@50$]{
    \begin{minipage}[t]{0.5\linewidth}
    \centering
    \includegraphics[scale=0.33]{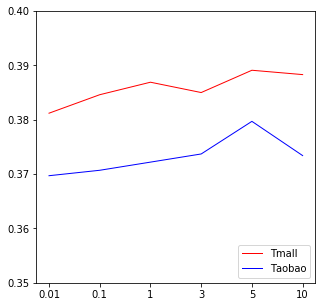}
    \end{minipage}}%
    \subfigure[Effect of $m^*$ on F$_1@50$]{
    \begin{minipage}[t]{0.5\linewidth}
    \centering
    \includegraphics[scale=0.33]{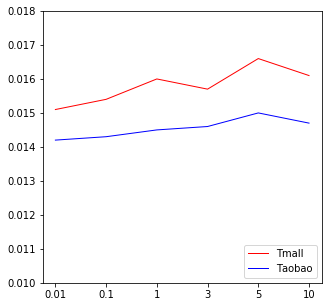}
    \end{minipage}}
    \centering
    \caption{The effect of margin parameter $m^*$.}
    % Since indexes are not in the same order of magnitude, and we only focus on the result fluctuation caused by the parameter change, so data are processed by min-max normalization.
    \label{fig:param2}
\end{figure}

The threshold $m^*$ is a parameter for distance control between clicked and unclicked sequence representation in a certain range, so that the unclicked representation has differentiation with random items. 
It also fits the hinge function to avoid correcting ``already correct” triplets within the threshold. 
A comparison experiment is performed on changes caused by $m^*$. 
% Parameter experiments follow the rule of coarse to fine \ie $\{0.01,0.1,1,3,5,10\}$. 
Figure~\ref{fig:param2} shows the change of the parameters $m^*$. %, which is taken from a finer set $\{0.01,0.1,1,3,5,10\}$. 
Result is the best when the parameter $m^*$ is 5, and it performs worse if $m^*$ is too large or too small. 
Similar observations could be drawn for MRR@50 and R@50, thus omitted due to space limitation. 
% The non-extreme dislike tendency of negative feedback becomes obvious.

\subsection{Online A/B Test}
We further conducted experiments on a much larger online dataset collected by Mobile Taobao App~(recommendation logs within one week), which contains about \textit{4 billion} user behavior sequences, \textit{30 million} high-quality items, and \textit{150 million} users.
Distributed training was executed over 20 parameter servers and 100 workers (P100 GPU with 16GB memory) considering the scalability of models. 
We kept the other parameters the same as offline experiments in the Section~\ref{sec:settings} and the training steps took more than 30 hours.
% After finishing the offline training, we further deployed the models into the production environment of two recommendation scenarios at Taobao, \textit{Guess you like} on the pages of \textit{shopping cart} and \textit{item detail} to test the performance of trained larger models.
We conducted the online A/B test for several weeks between our proposed XDM and SDM~\cite{lv2019sdm}, which was the previous deployed EBR model at Taobao.
We used a fast nearest neighbor embedding retrieval method from Equation~\ref{eq:ip}, to retrieve Top-$k$ items from the large-scale item pool.
The detailed deployment architecture followed SDM and we compared the same evaluation metric pCTR~(the Click-Through-Rate per page view where each page can recommend 20 items for a user).
The results show that XDM improves \textbf{3\%$\sim$4\%} averagely compared to SDM, which demonstrates the advantages of incorporating unclicked behavior sequences and our proposed method.  
Moreover, XDM has been successfully deployed on EBR system of several recommendation scenarios at Taobao since April, 2020.

\section{CONCLUSION}
In this paper, we study users' unclicked sequence modeling in sequential recommender in order to enrich user representations. 
The importance of unclicked items is emphasized and then incorporated into our new recommendation model. 
For modeling sequential behaviors with unclicked data, we design a novel model XDM, which adopts the symmetric metric learning with a triplet structure as well as confidence fusion network. 
The experiment results demonstrate the effectiveness of the proposed XDM and verify the importance of unclicked sequences in the sequential recommendation.
XDM has been fully deployed on EBR system of recommendation at Taobao.
% \section{Appendix}
For sake of the space, the appendix is provided in the external link: \href{https://github.com/alicogintel/XDM}{https://github.com/alicogintel/XDM}

\bibliographystyle{splncs04}
\bibliography{mybibliography}

\begin{thebibliography}{10}
\providecommand{\url}[1]{\texttt{#1}}
\providecommand{\urlprefix}{URL }
\providecommand{\doi}[1]{https://doi.org/#1}

\bibitem{covington2016deep}
Covington, P., Adams, J., Sargin, E.: Deep neural networks for youtube
  recommendations. In: RecSys. pp. 191--198 (2016)

\bibitem{ding2019reinforced}
Ding, J., Quan, Y., He, X., Li, Y., Jin, D.: Reinforced negative sampling for
  recommendation with exposure data. In: IJCAI. pp. 2230--2236 (2019)

\bibitem{gao2019learning}
Gao, C., He, X., Gan, D., Chen, X., Feng, F., Li, Y., Chua, T.S., Yao, L.,
  Song, Y., Jin, D.: Learning to recommend with multiple cascading behaviors.
  TKDE  (2019)

\bibitem{Hidasi:GRU4REC}
Hidasi, B., Karatzoglou, A., Baltrunas, L., Tikk, D.: Session-based
  recommendations with recurrent neural networks. arXiv preprint
  arXiv:1511.06939  (2015)

\bibitem{jean2014using}
Jean, S., Cho, K., Memisevic, R., Bengio, Y.: On using very large target
  vocabulary for neural machine translation. arXiv preprint arXiv:1412.2007
  (2014)

\bibitem{li2019multi}
Li, C., Liu, Z., Wu, M., Xu, Y., Zhao, H., Huang, P., Kang, G., Chen, Q., Li,
  W., Lee, D.L.: Multi-interest network with dynamic routing for recommendation
  at tmall. In: CIKM. pp. 2615--2623 (2019)

\bibitem{Li:NARM}
Li, J., Ren, P., Chen, Z., Ren, Z., Lian, T., Ma, J.: Neural attentive
  session-based recommendation. In: CIKM. pp. 1419--1428 (2017)

\bibitem{li2018learning}
Li, Z., Zhao, H., Liu, Q., Huang, Z., Mei, T., Chen, E.: Learning from history
  and present: Next-item recommendation via discriminatively exploiting user
  behaviors. In: KDD. pp. 1734--1743 (2018)

\bibitem{lv2019sdm}
Lv, F., Jin, T., Yu, C., Sun, F., Lin, Q., Yang, K., Ng, W.: Sdm: Sequential
  deep matching model for online large-scale recommender system. In: CIKM. pp.
  2635--2643 (2019)

\bibitem{pazzani2007content}
Pazzani, M.J., Billsus, D.: Content-based recommendation systems. In: The
  adaptive web, pp. 325--341. Springer (2007)

\bibitem{sarwar2001item}
Sarwar, B., Karypis, G., Konstan, J., Riedl, J.: Item-based collaborative
  filtering recommendation algorithms. In: WWW. pp. 285--295 (2001)

\bibitem{tang2018personalized}
Tang, J., Wang, K.: Personalized top-n sequential recommendation via
  convolutional sequence embedding. In: WSDM. pp. 565--573 (2018)

\bibitem{ijcai2020-349}
Xie, R., Ling, C., Wang, Y., Wang, R., Xia, F., Lin, L.: Deep feedback network
  for recommendation. In: IJCAI. pp. 2519--2525 (2020)

\bibitem{ying2018sequential}
Ying, H., Zhuang, F., Zhang, F., Liu, Y., Xu, G., Xie, X., Xiong, H., Wu, J.:
  Sequential recommender system based on hierarchical attention networks. In:
  IJCAI (2018)

\bibitem{zhao2018recommendations}
Zhao, X., Zhang, L., Ding, Z., Xia, L., Tang, J., Yin, D.: Recommendations with
  negative feedback via pairwise deep reinforcement learning. In: KDD. pp.
  1040--1048 (2018)

\end{thebibliography}




% LNCS Homepage, \url{http://www.springer.com/lncs}. Last accessed 4
% \end{thebibliography}
\end{document}